\documentclass[prb,preprint]{revtex4-2}
\usepackage[intlimits]{amsmath}
\usepackage{amsfonts}

\usepackage{enumitem}
\usepackage{psfrag}

\usepackage{amsthm}
\usepackage{float}
\usepackage{tikz}

\usepackage{ulem}

\newcommand{\strike}[1]{}

\newcommand\bea           {\begin{equation}\begin{array}l\displaystyle}

\usepackage[utf8]{inputenc}

\newcommand\beq            {\begin{equation}}
\newcommand\eeq            {\end{equation}}
\newcommand\eeqq            {\end{equation}}
\newcommand\bes           {\begin{subequations}}
\newcommand\esu           {\end{subequations}}

\renewcommand{\(}{\left(}
\renewcommand{\)}{\right)}
\renewcommand{\[}{\left[}
\renewcommand{\]}{\right]}

\newcommand{\bigx}[1]{\bBigg@{#1}}

\newcommand\p            {\partial}

\renewcommand\vec[1]{{\boldsymbol{#1}}}

\def\3pt#1#2#3{{\langle{#1}\vert{#2}\vert{#3}\rangle}}

\newcommand{\EQ}{\begin{equation}}
\newcommand{\EN}{\end{equation}}
\usepackage{epsfig}
\usepackage{psfrag}
\usepackage{amsmath}
\usepackage{graphicx}
\usepackage{amsfonts}
\usepackage{amssymb}
\usepackage{bm}

\def\tilde{\widetilde}

\def\hat{\widehat}
\def\*{\star}
\def\[{\left[}
\def\]{\right]}
\def\({\left(}      

\def\){\right)}

\def\frac#1#2{\dfrac{#1}{#2}}

\def\2pi{\hbox{$2\pi i$}}

\def\dsl{\raise.15ex\hbox{/}\kern-.57em\partial}
\def\Dsl{\,\raise.15ex\hbox{/}\mkern-.13.5mu D}

\def\2pi{\hbox{$2\pi i$}}

\def\dsl{\raise.15ex\hbox{/}\kern-.57em\partial}
\def\Dsl{\,\raise.15ex\hbox{/}\mkern-.13.5mu D}

\def\barray{\begin{eqnarray}}
\def\earray{\end{eqnarray}}
\def\beq{\begin{equation}}
\def\eeqq{\end{equation}}

\def\AA{\leavevmode\setbox0=\hbox{h}
\dimen0=\ht0 \advance\dimen0 by-1ex\rlap{\raise.67\dimen0\hbox{\char'27}}A}

\begin{document}

\bibliographystyle{plainnat}

\title{Free Fall of a Quantum Many-Body System}

\author{A. Colcelli}
\affiliation{SISSA and INFN, Sezione di Trieste, Via Bonomea 265, I-34136 
Trieste, Italy}

\author{G. Mussardo}
\affiliation{SISSA and INFN, Sezione di Trieste, Via Bonomea 265, I-34136 
Trieste, Italy}

\author{G. Sierra}
\affiliation{Instituto de F\'isica Te\'orica, UAM/CSIC, Universidad 
Aut\'onoma de Madrid, Madrid, Spain}

\author{A. Trombettoni}
\affiliation{Department of Physics, University of Trieste, Strada
  Costiera 11, I-34151 Trieste, Italy}
\affiliation{SISSA and INFN, Sezione di Trieste, Via Bonomea 265, I-34136
Trieste, Italy}

\date{\today}

\begin{abstract}

  The quantum version of the free fall problem is a topic often skipped in undergraduate quantum mechanics courses because its discussion usually requires wavepackets built on the Airy functions -- a difficult computation.
  Here, on the contrary, we show that the problem can be nicely simplified
  both for a single particle and for general many-body systems by
  making use of a gauge transformation that corresponds to
  a change of reference frame from the laboratory frame to the one comoving
  with the falling system. Using this approach, the quantum mechanics problem of a particle in an external gravitational potential
  reduces to a much simpler
  one where there is no longer any gravitational potential in the Schr\"{o}dinger equation.
  It is instructive to see that the same procedure can be used for many-body systems
  subjected to an external gravitational potential and a two-body interparticle potential that is a function of the distance between the particles.
  This topic provides a helpful and pedagogical example of a quantum many-body system
  whose dynamics can be analytically described in simple terms.  
\end{abstract}

\maketitle
\section{Introduction}

In classical mechanics, one of the first problems that students
encounter is the dynamics of a falling body: an object pulled down to the ground (\textit{e.g.} from  Pisa's tower)
by the constant force of Earth's gravity.
However, amazingly enough, the same problem is not always discussed in quantum mechanics courses due to the sharp contrast between 
the physical simplicity of the problem and the difficulty of its mathematical description. Basic quantum mechanics courses are largely structured around solving the time-dependent Schr\"odinger equation
$i \hbar \frac{\partial \psi}{\partial t}=H \psi$ for the wavefunction
$\psi(x,t)$ in terms of the eigenfunctions $\Psi$ which solve the time-independent equation $H\Psi=E\Psi$.
Indeed, in traditional approaches to the problem of determining the
wavefunction at time $t$, it is necessary to involve the Airy functions
and the projection of the falling body's wavefunction into this set of eigenfunctions. Resorting to the Eherenfest theorem provides expressions for position or momentum expectation values, but does not provide immediate insight on the simple solution of the system. Solving the problem using a time-dependent variational approach might reveal the solution's simple structure, but this technique is not introduced in many university quantum mechanics courses.
Here we 
show that an alternative way to deal with the quantum falling body is pedagogically simple , but also general enough to be applicable to the single particle case and to quantum many-body systems. This 
approach exploits the use of a gauge transformation of the wavefunction that corresponds to a change of reference frame from the inertial laboratory frame to the falling body's accelerated frame. By gauge transformation we mean the multiplication of the wavefunction by a phase factor. This multiplication will not affect expectation values of physical observables like the position of a wavepacket. In the new reference frame, there is of course no longer any gravitational effect and therefore the system appears to be {\it ``free"}, \textit{i.e.} not subject to gravity.  A few comments on terminology are in order.   Throughout the paper we sometimes refer to a system in the absence of gravity as ``free''.  This is not to be confused with an alternative meaning of ``free'' as ``non-interacting''.  Later, we will discuss interacting systems of particles; to avoid confusion we will refer to the interacting system in the absence of gravity as the ``non-falling'' system.  (And, of course, a system in ``free fall'' is not ``free'' in either sense of the word discussed above).  

It is worth emphasizing that the method discussed here can be applied to study the effect of the gravitational force on a quantum many-body
system where particles with position vectors $\vec{r}_j$ and $\vec{r}_k$ interact via a generic two-body potential of the form $V(\left| \vec{r}_j- \vec{r}_k \right|)$. This 
leads to some interesting results. For instance, as we discuss in the
following, the time evolution of observables, such as the variance of the position of a falling wavepacket, is the same as the time evolution for a free wavepacket. The effect of gravity shows up solely in the behavior of the expectation values of position (and powers thereof) which, on the other hand, can be obtained from the classical Newton's second law of motion. This last point follows from the Ehrenfest theorem (see
\textit{e.g.}~\cite{Griffiths}), from which we can infer that the momentum of the wavepacket varies linearly with time while its position has a quadratic time dependence. This last fact is valid for a generic interaction potential in any number of dimensions; in this paper, we will focus on one-
and three-dimensional cases as explanatory examples. We will also show how to easily determine the expressions for the energy and the total momentum of the falling many-body system
using the basic commutation rules. Finally, we show how to relate the one-body density matrix
of the falling body to the corresponding density matrix of the ``free" (although possibly interacting) non-falling system and give a simple relationship between the eigenvalues of the two density matrices.

Employing a gauge transformation to deal with quantum free fall was already presented in~\cite{Vandegrieft, Nauenberg} for the single particle case in one dimension only; here we will show that it can be extended to the case of interacting many-body quantum systems in more than one dimension, broadening the interest on this topic. In all of these examples we will work in the Schr\"odinger picture, in such a way that the method used to tackle the single quantum particle case can then be applied to the many-body problem, keeping the same formalism and hence giving a fluid extension of applicability.

\section{Free fall of a quantum particle} 

\begin{figure}[t]
\includegraphics[width=0.5\columnwidth]{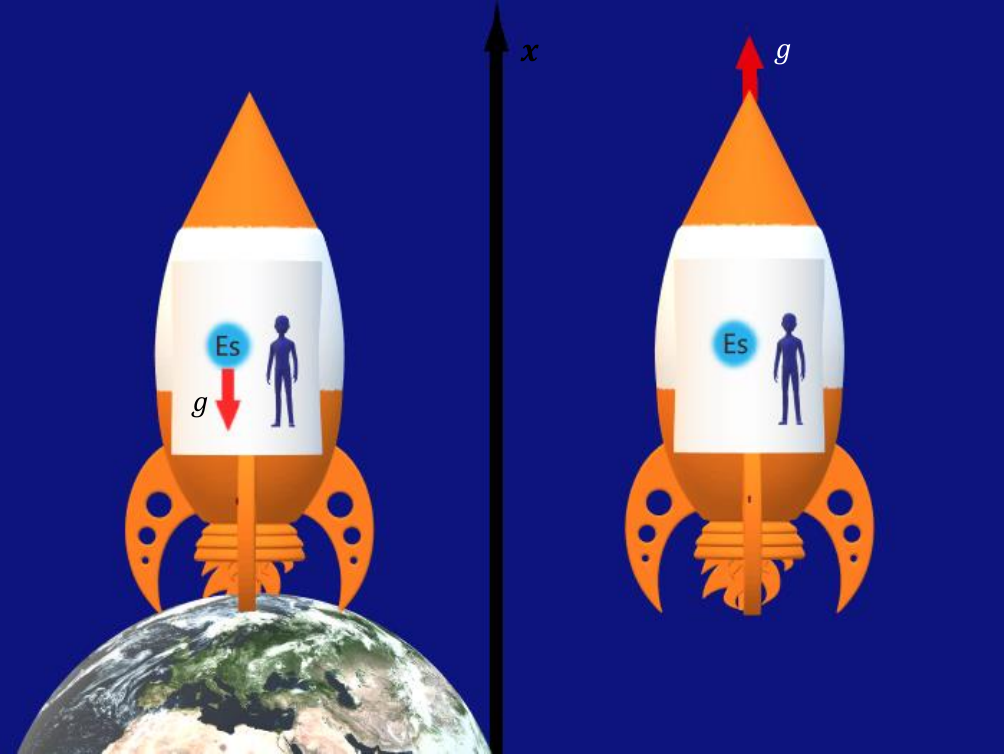}
\caption{Pictorial visualization of the Einstein's \textit{gedankenexperiment} of an Einsteinium atom inside an accelerating rocket  (right side picture) and a rocket at rest on Earth (left side picture).}
\label{fig1}
\end{figure}

We are interested in a quantum particle of mass $m$
subject to a linear gravitational potential (or, in a charged quantum particle subject to a constant electric field) in one dimension (see \textit{e.g.}~\cite{Nauenberg,LandauLifshitz}).  The case of a charged particle in a constant electric field is a straightforward generalization. The Schr\"odinger equation takes the form:
\beq
\label{schro_onebody}
i \hbar \frac{\p}{\p t}\psi(x,t)\,=\,\left(-\frac{\hbar^2}{2m} \frac{\p^2}{\p x^2} +m\, g\,x\right)\psi(x,t)\,,
\eeq
where $\psi(x,t)$ is the wave function and $g$ denotes the gravitational acceleration (for the case of a charged particle in a constant electric field $E$, one can replace $mg \to q E$). This equation may describe for example Einstein's famous gedankenexperiment of a rocket in empty space (i.e. very far from any other celestial body), and subject to an acceleration equal to $g \simeq 9.81 \frac{\rm m}{\rm s^2}$, as sketched in Fig. \ref{fig1}. Suppose that inside the rocket there is a single quantum object, e.g. an Einsteinium atom, for which the relevant Schr\"odinger equation will be indeed Eq. (\ref{schro_onebody}). The equation is also appropriate for the case when the rocket is standing still on the Earth's surface. 

This problem can be solved using the Fourier
transform to go to momentum space, as discussed in~\cite{Wadati1999}, while
other methods of solution, related to the Airy functions, were proposed and discussed in~\cite{Berry1978,Rau1996,Guedes2001,Feng2001}. For the solution of the problem involving a time dependent gravitational force see the recent paper~\cite{ourpaper}.

Given the initial wavepacket $\psi(x,t=0)$ subjected to a uniform gravitational field, what is the wavefunction $\psi(x,t)$ at a generic time $t$? As explained in the introduction, the solution that consists of projecting the initial wavepacket into the Airy functions is cumbersome, and other tools taught in basic quantum mechanics courses, such as time-dependent perturbation theory, may not prove illuminating. An easily accessible solution can be found by performing a gauge transformation which we now introduce.

\vspace{3mm}
\noindent
    {\bf Gauge Transformation}. Let us discuss now the method of solving Eq.~(\ref{schro_onebody}) by
    means of a gauge transformation. This result and approach is the same as~\cite{Nauenberg, Vandegrieft, Shukla, Kuchar}, but we review the derivation here in order to prepare for the presentation of the general case of a many-body system falling in three dimensions.  First, we pass to the comoving accelerated frame of reference by changing the spatial variable from $x$ to $\rho$, where 
\beq
\label{traslazione}
\rho(t)\,=\,x - \xi(t)
\eeqq
and $\xi(t)$ is related to the gravitational force acting on the system, as we are going to show. After this, we perform a gauge transformation\cite{footnote} via a space-time dependent phase $\theta(x,t)$, by writing the transformed wavefunction as
\beq
\label{wavefunction_gaugetrasf}
\psi(x,t) \,\equiv\, e^{i \theta(x,t)} \,\eta(\rho(t), t)\,,
\eeqq
with $\theta(x,t)$ a function to be determined in order to wash out the gravitational potential.

Substituting (\ref{wavefunction_gaugetrasf}) into (\ref{schro_onebody}),
we see that the external potential term can be eliminated if we impose
\beq
\label{conditions_integrab}
\frac{d\xi}{dt} \,=\,  \frac{\hbar}{m} \frac{\p \theta}{\p x} \,,
\,\,\,\,\,\,\,\, \,\,\,\,\,\,\,\,\,\,\,\,\, -\hbar \frac{\p \theta}{\p t} \,=\, \frac{\hbar^2}{2m} \left(\frac{\p \theta}{\p x}\right)^2 + m\,g\,x\,.
\eeqq
Assuming the validity of these equations, it is easy to see that $\eta(\rho,t)$ satisfies
the Schr\"odinger equation with no external potential but with $x$ replaced by the new spatial variable,$\rho$, and $\psi$ replaced by $\eta$,
\textit{i.e.}
\beq
\label{onebody_schro_eta}
i \hbar \frac{\p}{\p t}\eta(\rho,t)\,=\, -\frac{\hbar^2}{2\,m} \frac{\p^2 }{\p \rho^2}\eta(\rho,t)\,.
\eeqq
We observe that using Eq. (\ref{traslazione}) and employing Heisenberg representation would be an alternative way of solving the problem. We prefer to stick to the use of the gauge transformation in the Schr\"odinger picture because we believe that the method is instructive and useful for other calculations where gauge transformations are performed.

If we now make the ansatz 
\beq
\label{ansatz_theta}
\theta(x,t)\, =\, \frac{m}{\hbar} \frac{d\xi}{dt} \,x+ \Gamma(t)\,,
\eeqq
and use it in Eq.~(\ref{conditions_integrab}), we arrive at the conditions 
\beq
\label{vANDdelta_onebody}
m\frac{d^2 \xi}{dt^2} \,=\, -m\,g\,, \,\,\,\,\,\,\,\,\,\, \,\,\,\,\,\,\,\,\,\,\,\,\,\,\,\,\, \hbar \frac{d \Gamma}{dt}\,=\,-\frac{m}{2} \left(\frac{d\xi}{dt} \right)^2 \, , 
\eeqq
which determine the functions $\xi(t)$ and $\Gamma(t)$
in terms of the gravitational acceleration $g$. 
Once we solve the differential equations (\ref{vANDdelta_onebody}) with the initial conditions 
$\xi(0)= \dot{\xi}(0) = 0$ and $\Gamma(0)=0$, we get the following expression for the gauge phase
\beq
\label{theta_g}
\theta(x,t)\,=\,-\frac{m\,g\,t}{\hbar} \,x -\frac{m\,g^2 \,t^3}{6\,\hbar} \,,
\eeqq
while the  ``translational'' parameter $\xi$ reads
\beq
\label{xi_g}
\xi(t)\,=\, -\frac{g \,t^2}{2}\,\,\,.
\eeqq
Similar results may be obtained in the case $\dot{\xi}(0) \neq 0$.  Notice that the $x$-dependent term of the gauge phase in Eq. (\ref{theta_g}) is the gravitationally-induced phase difference that is observed in interferometers when two beams of particles follow paths that are at different heights. For a detailed pedagogical discussion about this, we refer readers to chapter $2.7$ of~\cite{SakuraiiNapolitano}. 

Eqs.~(\ref{theta_g}) and (\ref{xi_g}), together with Eqs.~(\ref{wavefunction_gaugetrasf}) 
and (\ref{onebody_schro_eta}), completely solve the Schr\"{o}dinger equation (\ref{schro_onebody}), 
since $\eta(\rho,t)$ is simply the well known solution of the Schr\"odinger equation for a free particle, which is often studied in quantum mechanics courses, for example in the context of the spreading of a Gaussian wavepacket. Notice that, with our choices $\theta(x,0)=0$ and 
$\rho(0)=x$, from (\ref{wavefunction_gaugetrasf}) we have $\psi(x,0)\,=\,\eta(x,0)$.
Therefore we can write the complete solution of the Schr\"odinger equation (\ref{schro_onebody}) as
\beq
\label{complete_sol_onebody}
\psi(x,t)\,=\,\exp\left[ i\theta(x,t)\right] \cdot \exp\left[-i \frac{t}{\hbar}\frac{\hat{p}^2}{2m}\right]\,\eta(\rho,0)\,=\,\exp\left[i\theta(x,t)\right]\cdot \exp\left[-i\frac{t}{\hbar}\frac{\hat{p}^2}{2m} -i \frac{\xi(t)}{\hbar} \hat{p}\right] \,\psi(x,0)\,,
\eeqq
where we used the definition of the translation operator
\beq
\label{def_transl_op}
\psi(x-\xi(t),t)\,=\,\exp\left[-i \frac{\xi(t)}{\hbar} \hat{p}\right]\,\psi(x,t)\,,
\eeqq
and the free time evolution operator. In Eqs.~(\ref{complete_sol_onebody}) and
(\ref{def_transl_op}), $\hat{p}$ refers to the momentum operator: $\hat{p}\rightarrow-i\hbar\frac{\p}{\p x}$. 

\vspace{3mm}
\noindent
    {\bf Expectation values}. Using the results just discussed we can study
    the time evolution of expectation values of different operators, such as position and momentum. The expectation values of powers of $\hat{x}$ are defined as
\beq
\left\langle x^{\mathcal{N}}\right\rangle(t)\,\equiv\,\left \langle \psi(x,t)|\,\hat{x}^{\mathcal{N}}\,| \psi(x,t)\right \rangle \,=\,\int_{-\infty}^{\infty} \left|\psi(x,t)\right|^2 \,x^{\mathcal{N}}\,dx\,,\
\eeqq
while the expectation values of powers of the momentum $\hat{p}$ are
\beq
\left\langle p^{\mathcal{N}}\right\rangle(t)\,\equiv\,\left \langle \psi(x,t)| \,\hat{p}^{\mathcal{N}}\, | \psi(x,t)\right \rangle \,=\,(-i\hbar)^{\mathcal{N}} \int_{-\infty}^{\infty} \psi^*(x,t) \,\frac{\p^{\mathcal{N}}}{\p x^{\mathcal{N}}} \psi(x,t)\,dx\, 
\eeqq
(where the wavefunction $\psi(x,t)$ is normalized).

Assuming initial values
\beq
\label{initial_conditions_xp}
\left \langle x\right \rangle (0)\,=\,x_0\,, \,\,\,\,\,\,\,\,\,\,\,\,\,\,\,\,\,\,\,\, \left \langle p\right \rangle (0)\,=\,p_0\,,
\eeqq
we can employ the solution (\ref{complete_sol_onebody}) to obtain the time evolved expectation
values. In the following, we focus our attention on the
cases ${\mathcal{N}}=1,2$ since these quantities are commonly encountered in 
introductory quantum mechanics courses.

\vspace{3mm}
\noindent
{\bf Commutation relations}.
Before proceeding in that direction, it is useful to study commutation relations
among different operators, such as the position operator with the translation and
the time evolution operators. A similar insightful discussion on operator algebra in the case of gravitational problems can be found in~\cite{WheelerNotes}.

Writing 
$$\theta(\hat{x},t)\equiv \hat{x} A+B\,,$$
where in our case 
$$A \,=\,-\frac{m\,g\,t}{\hbar}
\,, \,\,\,\,\,\,\,\,\,\,\,\,\,\,\,
B=-\frac{m\,g^2 \,t^3}{6\,\hbar}\,\,\,,
$$
we have
\beq
\label{comm_rel_5}
\left[\hat{p},e^{i\,\theta(\hat{x},t)}\right]\,=\,\hbar\,A\,e^{i\,\theta(\hat{x},t)}\,.
\eeq
Finally we need to calculate
\begin{eqnarray}
\nonumber
\left[\hat{p}^2,e^{i\,\theta(\hat{x},t)}\right]&\,=\,&\hat{p}\,\left[\hat{p},e^{i\,\theta(\hat{x},t)}\right]+\left[\hat{p},e^{i\,\theta(\hat{x},t)}\right]\,\hat{p}\,=\,\hbar\,A\,\hat{p}\,e^{i\,\theta(\hat{x},t)}+\hbar\,A\,e^{i\,\theta(\hat{x},t)}\hat{p}\,=\,
\\ \label{comm_rel_6}
&\,=\,& \left(\hbar\,A\right)^2\,e^{i\theta(\hat{x},t)}+2\,\hbar\,A\,e^{i\theta(\hat{x},t)}\hat{p}\,,
\end{eqnarray}
where we used the result (\ref{comm_rel_5}) . Similar results for the commutation relation of the position  operator (and its powers) may be easily found. 

\vspace{3mm}
\noindent
{\bf Time evolution of operators involving ${\bf \hat x}$}. 
We have all quantities we need to evaluate expectation values of the state $\psi(x,t)$
in (\ref{complete_sol_onebody}). Let us start with 
\beq
\left \langle \psi(x,t)|\,\hat{x}\,| \psi(x,t)\right \rangle \,=\,\left \langle \eta(\rho,0\,)\Big|\,
\exp\left[i\frac{t}{2\,m\,\hbar}\,\hat{p}^2 \right] \,\hat{x}\, \exp\left[- i\frac{t}{2\,m\,\hbar}\hat{p}^2\right]\,\Big| \,\eta(\rho,0)\right \rangle\,,
\eeq
where we used the fact that $\hat{x}$ commutes with $e^{i\,\theta(\hat{x},t)}$.
With the help of the commutation relations $\left[\hat{x},\hat{p}\right]=i\hbar$, and 
\begin{eqnarray}
\left[\hat{x},e^{-i\,a\,\hat{p}} \right] &=& \hbar\,a\,e^{-i\,a\,\hat{p}} \\
\left[\hat{x}^2,e^{-i\,a\,\hat{p}}\right] &=& \hbar \, a\, e^{-i\,a\,\hat{p}}\,\left(\hbar\,a\,e^{-i\,a\,\hat{p}}+2\,\hat{x}\right) \\ 
\left[\hat{x},e^{-i\,b\,\hat{p}^2} \right] &=& 2\,\hbar\,b\,e^{-i\,b\,\hat{p}^2}\hat{p} \\
\left[\hat{x}^2,e^{-i\,b\,\hat{p}^2} \right] &=& 4\,\hbar\,b\,e^{-i\,b\,\hat{p}^2}\hat{p}\,\hat{x}+(2\,\hbar\,b)^2\,e^{-i\,b\,\hat{p}^2}\hat{p}^2+2\,i\,\hbar^2\,b\,e^{-i\,b\,\hat{p}^2} 
\end{eqnarray}
where $a$ and $b$ are generic real parameters, we get
\begin{eqnarray}
\label{x_result}
\left \langle \psi(x,t)|\,\hat{x}\,| \psi(x,t)\right \rangle &=& \frac{t}{m}\,p_0+\xi(t)\left \langle \eta(x,0)| \eta(x,0)\right \rangle+\left \langle \eta(x,0)|\,\hat{x}\,| \eta(x,0)\right \rangle \\ 
	&=&\,\frac{t}{m}\,p_0+\xi(t)+x_0 \, = \xi(t) + \langle x \rangle_{\rm free}(t),
\end{eqnarray}
and
\beq
\label{x^2_result}
\left \langle \psi(x,t)\big|\,\hat{x}^2\,\big| \psi(x,t)\right \rangle \,=\,\xi^2(t)+2\,\xi(t)\frac{t}{m}\,p_0+2\,\xi(t)\,x_0+\left\langle x^2\right\rangle_{\rm free}(t)\,,
\eeq
where we have employed the normalization condition and we have defined
\begin{eqnarray}
\left\langle x \right\rangle_{\rm free}(t) \, &\equiv& \left \langle \eta(x,t) \big| \, \hat{x} \, \big| \eta(x,t) \right \rangle \\
\left\langle x^2\right\rangle_{\rm free}(t) &\equiv& \left \langle \eta(x,t)\big|\,\hat{x}^2\,\big| \eta(x,t)\right \rangle\,,
\end{eqnarray}
which are expectation values of $x^2$ evaluated on the ``free" Schr\"odinger
equation solution $\eta(x,t)$ \textit{i.e.} with $g=0$, prepared in the initial state $\eta(x,0)=\psi(x,0)$. 

Finally, 
we get the following expression for the standard deviation
\beq
\label{variancex_result}
\Delta x(t)\,=\,\sqrt{\left\langle x^2\right\rangle_{\rm free}(t)-\left(\frac{t}{m}\,p_0+x_0\right)^2}\,=\,\sqrt{\left\langle x^2\right\rangle_{\rm free}(t)-\left[\left\langle x\right\rangle_{\rm free}(t)\right]^2}\,\equiv\,\Delta x_{\rm free}(t)\,.
\eeq
As evident from Eq.~(\ref{variancex_result}), the variance of the position of the
falling wavepacket behaves the same as in the free (\textit{i.e.} no gravity) case. 

\vspace{3mm}
\noindent
    {\bf Time evolution of ${\bf \hat p}$'s operators}. Similar computations
    can be done for the expectation values involving the momentum. 
Proceeding as before, we get
\beq
\label{p_result}
\left \langle \psi(x,t)|\,\hat{p}\,| \psi(x,t)\right \rangle\,=\,-m\,g\,t+p_0\,=\,-m\,g\,t+\left\langle p\right\rangle_{\rm free}(0)\,,
\eeq
and
\beq
\label{p^2_result}
\left \langle \psi(x,t)|\,\hat{p}^2\,| \psi(x,t)\right \rangle\,=\,m^2\,g^2\,t^2-2\,m\,g\,t\,p_0+\left\langle p^2\right\rangle_{\rm free}(0)\,,
\eeq
where we used the commutation relation (\ref{comm_rel_6}) with $A=-(m\,g\,t)/\hbar$,
and we defined 
$$
\left\langle p^2\right\rangle_{\rm free}(t)\equiv\left \langle \eta(x,t)|\,\hat{p}^2\,| \eta(x,t)\right \rangle=\left \langle \eta(x,0)|\,\hat{p}^2\,| \eta(x,0)\right \rangle\equiv\left\langle p^2\right\rangle_{\rm free}(0)\,\,\,.
$$
Finally, let's compute the standard deviation of $\hat{p}$ using Eqs.~(\ref{p_result}) and (\ref{p^2_result}):
\begin{eqnarray}
\label{variancep_result1}
\Delta p(t)\,=\,\sqrt{\left\langle p^2\right\rangle(t) - \left\langle p \right \rangle^2(t)}&\,=\,&\sqrt{\left\langle p^2\right\rangle(0) - (p_0)^2}\,\equiv\,\Delta p(0)\\ \label{variancep_result2}
&\,=\,&\sqrt{\left\langle p^2\right\rangle_{\rm free}(t)-\left[\left\langle p\right\rangle_{\rm free}(t)\right]^2}\,\equiv\,\Delta p_{\rm free}(t)\,.
\end{eqnarray}
According to (\ref{variancep_result1}), the variance of the momentum remains equal to its initial
value at $t=0$, while Eq.~(\ref{variancep_result2}) shows that the evolution of the variance of
$\hat{p}$ for a falling wavepacket is exactly the same as that of a freely expanding wavepacket.

%
We can check the results obtained above in the case of an initial Gaussian wavepacket subject to a gravitational force, reproducing the results of~\cite{Nauenberg}.
We prepare a Gaussian wavepacket centered on $x_0$ with standard deviation $\sigma$, and
    with initial momentum $k_0$:
\beq
\label{initial_GWP}
\psi(x,0)\,=\,\frac{1}{\sqrt[4]{2\pi\sigma^2}}\,\exp\left[i\,k_0\,x -\frac{(x-x_0)^2}{4\,\sigma^2}\right]\,.
\eeq
One gets
\begin{eqnarray}
\nonumber
\psi(x,t)\,&=&\,\frac{1}{\sqrt[4]{2\,\pi\,\sigma^2}}\,\frac{e^{i\,\theta(x,t)}}{\sqrt{1+i\,\frac{\hbar\,t}{2\,m\,\sigma^2}}} \\
\label{GWP_timeevolved} 
&& \times \exp{\left\{-\frac{\left[x-x_0-\xi(t)\right]^2+4\,i\,k_0\,\sigma^2\left[x-x_0-\xi(t)\right]+2\,i\,\hbar\,t\,(k_0\,\sigma)^2/m}{4\left(\sigma^2+i\,\frac{\hbar\,t}{2\,m}\right)}\right\}}\,,
\end{eqnarray}
with variance
\beq
\label{variance_gaussian}
\Delta x(t)\,=\,\sqrt{\sigma^2+\frac{\hbar^2\,t^2}{4\,m^2\,\sigma^2}}\,,
\eeq
while the motion of the centers of mass coincides with the expected value given in
Eq.~(\ref{x_result}).

\vspace{3mm}
\noindent
{\bf Three-dimensional case}. 
It is simple to extend the analysis which we presented above to the case of a single
particle falling along the x-direction in a three-dimensional space. In this case the Schr\"odinger equation reads
\beq
\label{3d_schro_1body}
i \hbar \frac{\p}{\p t}\psi(\vec{r},t)\,=\,i \hbar \frac{\p}{\p t}\psi(x,y,z,t)\,=\,\left(-\frac{\hbar^2}{2m} \vec{\nabla}_x^2 +m\, g\,x\right)\psi(\vec{r},t)\,,
\eeq
where the vector position $\vec{r}$ is expressed in Cartesian coordinates in the second equality,
and we have denoted the Laplacian
\beq
\nonumber
\vec{\nabla}^2_x\,\equiv\,\frac{\p^2}{\p x^2}+\frac{\p^2}{\p y^2}+\frac{\p^2}{\p z^2}\,.
\eeq
Proceeding in the same way as for the $1D$ case,
we perform a translation and a gauge transformation 
\beq
\label{gauge_trasf_3d_onebody}
\psi(\vec{r},t)\,=\,e^{i\,\theta(x,t)}\,\eta(\rho(t),y,z,t)\,,
\eeq
where $\rho(t)=x-\xi(t)$, and the gauge phase $\theta(x,t)$ and the translational parameter
$\xi(t)$ satisfy Eqs.~(\ref{conditions_integrab}). With these conditions,
the Schr\"odinger equation (\ref{3d_schro_1body}) is reduced to the free Schr\"odinger
equation for $\eta(\rho,y,z,t)$:
\beq
i \hbar \frac{\p}{\p t}\eta(\rho,y,z,t)\,=\,-\frac{\hbar^2}{2m} \vec{\nabla}_\rho^2\,\eta(\rho,y,z,t)\,,
\eeq
where we define
\beq
\nonumber
\vec{\nabla}^2_\rho\,\equiv\,\frac{\p^2}{\p \rho^2}+\frac{\p^2}{\p y^2}+\frac{\p^2}{\p z^2}\,.
\eeq
Analogously to (\ref{complete_sol_onebody}), choosing $\theta(x,t)$ to be (\ref{theta_g})
and $\xi(t)$ given by Eq.~(\ref{xi_g}), we can rewrite (\ref{gauge_trasf_3d_onebody}) as
\beq
\label{complete_sol_3d_onebody}
\psi(\vec{r},t)\,=\,\exp\left[i\theta(x,t)\right]\cdot\exp\left\{-i\frac{t}{\hbar}\frac{\hat{\vec{p}}\,^2}{2m} -i \frac{\xi(t)}{\hbar} \hat{p}_x\right\}\,\,\psi(\vec{r},0)\,,
\eeq
where we have defined: $\hat{\vec{p}}\,^2\,=\,\hat{p}_x^2+\hat{p}_y^2+\hat{p}_z^2$,
with the momentum operators defined as
$\hat{p}_\alpha\rightarrow -i\,\hbar\,\frac{\p}{\p \alpha}$ where $\alpha$ labels the $x$, $y$, or $z$ component. We are now able to study how
expectation values of different physical quantities evolve. First we redefine expectation
values of position operator and its powers as
\beq
\left\langle \alpha^{\mathcal{N}}\right\rangle(t)\,\equiv\,\left \langle \psi(\vec{r},t)|\,\hat{\alpha}^{\mathcal{N}}\,| \psi(\vec{r},t)\right \rangle \,=\,\int_{-\infty}^{\infty}\,dx\,\int_{-\infty}^{\infty}\,dy\,\int_{-\infty}^{\infty}\,dz\, \left|\psi(\vec{r},t)\right|^2 \,\alpha^{\mathcal{N}}\,,
\eeq
where, again, $\alpha$ stands for the $x$, $y$ or $z$ coordinate, while for the powers of the momentum
\beq
\left\langle p_\alpha^{\mathcal{N}}\right\rangle(t)\,\equiv\,\left \langle \psi(\vec{r},t)| \,\hat{p}_\alpha^{\mathcal{N}}\, | \psi(\vec{r},t)\right \rangle \,=\,(-i\hbar)^{\mathcal{N}} \int_{-\infty}^{\infty}\,dx\,\int_{-\infty}^{\infty}\,dy\,\int_{-\infty}^{\infty}\,dz\,\psi^*(\vec{r},t) \,\frac{\p^{\mathcal{N}}}{\p \alpha^{\mathcal{N}}} \psi(\vec{r},t)\,.
\eeq
From Eq.~(\ref{complete_sol_3d_onebody}) it is straightforward to perform
the calculation of expectation values of different coordinates, and since operators
associated with different coordinate axes commute (like $\hat{x}$ and $\hat{p}_y$ or $\hat{p}_y$ and
$\hat{p}_z$ and so on) then the motion in the $y$ and $z$ directions is trivially evaluated
to be the free one ($g=0$), while for the $x$ component one relies on the
results presented previously for the one-dimensional case. 

In summary, the motion of a wavepacket in three-dimensions under
the action of gravity is described by a spreading which is identical to a free (no gravity) expansion in all directions, while its center of mass accelerates in the direction of the gravitational force as a classical particle would.

\section{Free fall of two interacting particles}

In the previous section we revisited the solution for the quantum dynamics
  of a particle falling under the action of a gravitational potential. The formalism was presented in a way that can be straightforwardly extended to the case of two or more particles. Such results may be of interest in advanced courses, and may have other modern applications.

We start by studying a three-dimensional system of two interacting particles subject to gravity.
The Schr\"odinger equation reads 
\beq
\label{schro_twobodies}
i\hbar \frac{\p}{\p t} \psi(\vec{r}_1,\vec{r}_2,t)\,=\,\left[-\frac{\hbar^2}{2m}\left(\vec{\nabla}_{r_1}^2+\vec{\nabla}_{r_2}^2\right) +V(\left|\vec{r}_2-\vec{r}_1\right|)+mg \left(x_1+x_2\right) 
\right]\psi(\vec{r}_1,\vec{r}_2,t)\,,
\eeq
where
\beq
\label{nabla_definition}
\vec{\nabla}_{r_j}^2\,\equiv\,\frac{\p^2}{\p x_j^2}+\frac{\p^2}{\p y_j^2}+\frac{\p^2}{\p z_j^2}\,,
\eeq
for $j=1,2$, and $V(\left|\vec{r}_2-\vec{r}_1\right|)$ describes the interaction among particles and depends only on the distance between them.
In order to solve the Schr\"odinger equation, we employ the same method outlined in
the previous section: we perform a translation and a gauge transformation on the wavefunction
\beq
\label{sol_twobodies}
\psi(\vec{r}_1,\vec{r}_2,t)\,=\,e^{i\left[\theta(x_1,t)+\theta(x_2,t)\right]}\,
\eta(\boldsymbol{\varrho}_1(t),\boldsymbol{\varrho}_2(t),t)\,,
\eeq 
where 
we defined the vector
  $\boldsymbol{\varrho}_j(t)=\left(x_j-\xi(t)\,,\,y_j\,,\,z_j\right)$ and we set
$\rho_j (t) \equiv x_j-\xi(t)$ 
for $j=1,2$.  $\theta(x_1,t)$ obeys
Eq.~(\ref{conditions_integrab}) for $x=x_1$, while $\theta(x_2,t)$ obeys Eq.~(\ref{conditions_integrab}) for $x=x_2$. Notice that because
the interaction potential depends on distance between the particles, it remains unchanged as a result of the definition of the new spatial variables $\rho_j(t)$
and $\boldsymbol{\varrho}_j(t)$.
Using the ansatz (\ref{ansatz_theta}) and Eq.~(\ref{vANDdelta_onebody}), 
$\eta(\varrho_1,\varrho_2,t)$ will satisfy the free Schr\"odinger equation for two
interacting particles 
\beq
\label{twobodies_schro_eta}
i\hbar \frac{\p}{\p t} \eta(\boldsymbol{\varrho}_1,\boldsymbol{\varrho}_2,t)\,=\,\left[-\frac{\hbar^2}{2m}\left(\vec{\nabla}_{\varrho_1}^2+\vec{\nabla}_{\varrho_2}^2\right)  +V(\left|\boldsymbol{\varrho}_2-\boldsymbol{\varrho}_1\right|) \right]\eta(\boldsymbol{\varrho}_1,\boldsymbol{\varrho}_2,t)\,,
\eeq
with
\beq
\label{nabla_many_def}
\vec{\nabla}_{\varrho_j}^2\,\equiv\,\frac{\p^2}{\p \rho_j^2}+\frac{\p^2}{\p y_j^2}+\frac{\p^2}{\p z_j^2}\,.
\eeq
Therefore if one knows how to solve Eq.~(\ref{twobodies_schro_eta}),
then the complete solution of (\ref{schro_twobodies}) reads
\beq
\label{complete_sol_twobodies}
\psi(\vec{r}_1,\vec{r}_2,t)\,=\,\exp\left[-i\frac{m\,g\,t}{\hbar}\left(\frac{g\,t^2}{3}+x_1+x_2\right)\right]\,\eta\left(x_1+\frac{g\,t^2}{2},y_1,z_1 ; x_2+\frac{g\,t^2}{2}, y_2,z_2;t\right)\,.
\eeq

We can now ask the same questions as before: if we start from a generic wavepacket
$\psi(r_1,r_2,0)$ and we let it evolve under the action of gravity, how do its variances and expectation values of powers of position behave? Let's define as usual
\beq
\label{expect_value_pos_def}
\left\langle \alpha_j^{\mathcal{N}}\right\rangle(t)\,\equiv\,\left \langle \psi(\vec{r}_1,\vec{r}_2,t)|\,\hat{\alpha}_j^{\mathcal{N}}\,| \psi(\vec{r}_1,\vec{r}_2,t)\right \rangle \,=\,\int 
d^3r_1\,\int d^3r_2 \,\left|\psi(\vec{r}_1,\vec{r}_2,t)\right|^2 \,\alpha_j^{\mathcal{N}}
\eeq
where $\alpha$ can be either $x$, $y$ or $z$, while $j=1,2$ labels the particles.
For the expectation value of powers of the momenta $\hat{p}_{\alpha_j}$
we have 
\beq
\label{expect_value_mom_def}
\left\langle p_{\alpha_j}^{\mathcal{N}}\right\rangle(t)\,\equiv\,\left \langle \psi(\vec{r}_1,\vec{r}_2,t)|\,\hat{p}_{\alpha_j}^{\mathcal{N}}\,| \psi(\vec{r}_1,\vec{r}_2,t)\right \rangle 
\,=\,(-i\hbar)^{\mathcal{N}} \int\,d^3r_1\,\int\,d^3r_2\, \psi^*(\vec{r}_1,\vec{r}_2,t) \,\frac{\p^{\mathcal{N}}}{\p \alpha_j^{\mathcal{N}}} \psi(\vec{r}_1,\vec{r}_2,t)
\eeq
with $\int d^3r_j =\int_{-\infty}^\infty dx_j\,\int_{-\infty}^\infty dy_j\,\int_{-\infty}^\infty dz_j$. For
the initial conditions we take
\beq
\left \langle \alpha_j\right \rangle (0)\,=\,\alpha_0^{(j)}\,, \,\,\,\,\,\,\,\,\,\,\,\,\,\,\,\,\,\,\,\,
\left \langle p_{\alpha_j}\right \rangle (0) \,=\,p_{\alpha 0}^{(j)}\,.
\label{conditions_integrab_bis}
\eeq

It is actually very simple to show that the same results for the 
one-particle case will hold, that is to say the variances of positions of the particles
will behave as the free expanding case, while the expectation values of powers of the
$x$ component for positions have the same expressions of the one-body case, see
Eqs.~(\ref{x_result}) and (\ref{x^2_result}), with an additional index $j=1,2$ to
label the particles. For $y$ and $z$ components, one
has the formulas with $g=0$, since
the gravitational potential only affects motion in the $x$ direction.
The simplicity of this result comes from the fact that the commutators
among operators acting on different particles vanish,
therefore \cite{footnote2}
\begin{gather}
\nonumber
\left[\hat{x}_j,e^{-i\,a\,\hat{p}_{x_k}}\right]\,=\,\hbar\,a\,e^{-i\,a\,\hat{p}_{x_k}}\,\delta_{j,k}\,,\\ \nonumber
\left[\hat{x}_j^2,e^{-i\,a\,\hat{p}_{x_k}}\right]\,=\,\hbar \, a\, e^{-i\,a\,\hat{p}x_k}\,\left(\hbar\,a\,e^{-i\,a\,\hat{p}x_k}+2\,\hat{x}_j\right)\,\delta_{j,k}\,,\\ \nonumber
\left[\hat{x}_j,e^{-i\,b\,\hat{p}_{x_k}^2}\right]\,=\,2\,\hbar\,b\,e^{-i\,b\,\hat{p}_{x_k}^2}\,\hat{p}_{x_k}\,\delta_{j,k}\,,\\ \nonumber
\left[\hat{x}_j^2,e^{-i\,b\,\hat{p}_{x_k}^2}\right]\,=\,\left[4\,\hbar\,b\,e^{-i\,b\,\hat{p}_{x_k}^2}\hat{p}_{x_k}\,\hat{x}_j+(2\,\hbar\,b)^2\,e^{-i\,b\,\hat{p}_{x_k}^2}\hat{p}_{x_k}^2+2\,i\,\hbar^2\,b\,e^{-i\,b\,\hat{p}_{x_k}^2}\right]\delta_{j,k}\,,
\end{gather}
where $\delta_{j,k}$ is the Kronecker delta and $a$ and $b$ are again scalar quantities. 
We can rewrite (\ref{sol_twobodies}) as
\begin{eqnarray}
\psi(\vec{r}_1,\vec{r}_2,t)&\,=\,& 
\exp\left\{ i[\theta(\hat{x}_1,t)+\theta(\hat{x}_2,t)]\right\}\cdot\exp\left\{ - i\frac{t}{\hbar}\left[\frac{\hat{p}_1^2+\hat{p}_2^2}{2m} + 
V\left(\left|\hat{\varrho}_2-\hat{\varrho}_1\right|\right)\right]\right\}\,\eta(\varrho_1,\varrho_2,0) \nonumber \\
&\,=\,&\exp\left\{ i [\theta(\hat{x}_1,t)+\theta(\hat{x}_2,t)]\right\}\cdot\exp\left\{ - i\frac{t}{\hbar}\hat{H}_0 - i \frac{\xi(t)}{\hbar} (\hat{p}_{x_1}+\hat{p}_{x_2})\right\}\,\psi(\vec{r}_1,\vec{r}_2,0)\,,
\end{eqnarray}
where we have defined 
$$
\hat{H}_0\equiv \frac{\hat{p}_1^2 +\hat{p}_2^2}{2m} + V\left(\left|\hat{\varrho}_2-\hat{\varrho}_1\right|\right)\,\,\,.
$$ 
One can repeat the exact same steps performed in the previous section to obtain
the expectation values involving position and momentum. We summarize
the results below for the general case of two particles having different initial velocities according to
  Eqs.\eqref{conditions_integrab_bis}:
\begin{gather}
\label{1_result}
\left \langle x_j\right\rangle (t)\,=\,\frac{t}{m}\,p_{x0}^{(j)}+\xi(t)+x_0^{(j)}\,,\\
\label{2_result}
\left \langle x_j^2\right\rangle (t)\,=\,\xi^2(t)-2\,\xi(t)\,\frac{t}{m}\,p_{x0}^{(j)}-2\,\xi(t)\,x_0^{(j)}+\left\langle x_j^2\right\rangle_{\rm free}(t)\,,\\
\label{5_result}
\Delta x_j(t)\,=\,\sqrt{\left\langle x_j^2\right\rangle(t)-\left\langle x_j\right\rangle^2(t)}\,=\,\left(\Delta x_j\right)_{\rm free}(t)\,,
\end{gather}
where, as before, the subscript ``free'' denotes the expectation values in the freely expanding state $\eta(r_1, r_2,t)$ \cite{footnote3}.
The same expressions, but with $g=0$, are valid for the expectation values on the $y$ and $z$ components. 

We conclude that this method of gauge transformation can be used to reduce the initial
Schr\"odinger equation describing the dynamics of two falling interacting particles (with potentially different masses and velocities) to the
simpler Schr\"odinger equation where no gravitational potential is present, and with the
same interaction potential among the particles. The fundamental requirement is that the
two-body potential depends only on the relative distance between the particles.

\section{Free fall of a quantum many-body system}

In the literature, the motion of a ``structured", many-body
quantum system under the action of a gravitational potential has been studied in various situations ranging from Bose-Einstein condensates \cite{ChenLiu1976, Wadati2001, Ablowitz2004}
to one-dimensional integrable systems \cite{Sen1988, JukicGalic2010}. The case of a general three-dimensional many-body system subject to gravity can be explicitly addressed using the method described in the previous sections.

Let's then focus our attention on the Schr\"odinger equation for $N$ interacting particles (which are spinless, for simplicity) subject to a gravitational force along the $x$ direction 
\beq
\label{schro_Nbodies}
i\hbar \frac{\p}{\p t} \psi(\vec{r}_1,\dots,\vec{r}_N,t)\,=\,\left[-\frac{\hbar^2}{2m}\sum_{j=1}^N \vec{\nabla}_{r_j}^2 +\sum_{j<k} V(\left|\vec{r}_k - \vec{r}_j\right|)+m\,g \sum_{j=1}^N x_j \right]\,
\psi(\vec{r}_1,\dots,\vec{r}_N,t)\,,
\eeq
where the second sum is a double summation running over the two indicies $j$ and $k$, the interaction potential depends on the relative distances among particles, and the kinetic part is written in terms of (\ref{nabla_definition}).
In this more general case, the translation and gauge transformation take the form
\beq
\label{sol_Nbodies}
\psi(\vec{r}_1,\dots,\vec{r}_N,t)\,=\,\prod_{j=1}^N e^{i \theta(x_j,t)}\,\eta(\boldsymbol{\varrho}_1(t),\dots,\boldsymbol{\varrho}_N(t),t)\,,
\eeq 
which is a trivial generalization to the $N$ particle case of Eq.~(\ref{sol_twobodies}).
If the gauge phase $\theta(x_j,t)$ and the translational parameter $\xi(t)$ satisfy
(\ref{conditions_integrab}) with $x=x_j$, then $\eta(\varrho_1(t),\dots,\varrho_N(t),t)$
is the solution of the free Schr\"odinger equation
\beq
\label{Nbodies_schro_eta}
i\hbar \frac{\p}{\p t} \eta(\boldsymbol{\varrho}_1,\dots,\boldsymbol{\varrho}_N,t)\,=\,\left[-\frac{\hbar^2}{2m} \sum_{j=1}^N \vec{\nabla}_{\varrho_j}^2 +\sum_{j<k} V(\left|\boldsymbol{\varrho}_k-\boldsymbol{\varrho}_j\right|) \right]\eta(\boldsymbol{\varrho}_1,\dots,\boldsymbol{\varrho}_N,t)\,,
\eeq
where the kinetic part is expressed in terms of (\ref{nabla_many_def}) for every $j$.
Using Eq.~(\ref{sol_Nbodies}) one can prove that all results presented
in the previous sections, 
in particular Eqs.~(\ref{1_result}) -- (\ref{5_result}) and the
same expressions for $y$ and $z$ coordinates but with $g=0$, also hold for the many-body system. 

Writing $\vec{r}\,=\,x \, \hat{i}+y\,\hat{j}+z \, \hat{k}$, where $\hat{i}$, $\hat{j}$ and $\hat{k}$ are the usual unit vectors in Cartesian coordinate system, then we can write $\left \langle \vec{r}\right \rangle(t)\,=\, \hat{i}\, \left \langle x\right \rangle(t)+\hat{j} \, \left \langle y\right \rangle(t)+\hat{k}\, \left \langle z\right \rangle(t)$. Using this relation (and the analogous relation for the vector momentum $\vec{P}$), one can derive
the time evolution of expectation values of the wavepacket's position (and momentum). 
In particular, given that the system conserves the $x$, $y$ and $z$-components of the
total momentum, \textit{i.e.} $\left[\hat{H}_0, \sum_{j=1}^N \hat{p}_{\alpha_j} \right]=0$
for $\alpha=x$, $y$ and $z$, one can explicitly work out the total momentum and
the energy expectation values of the system in terms of the free case ($g=0$).
Using the commutation relations obtained previously, one gets that
\beq
\label{total_momentum}
\left \langle \vec{P} \right\rangle(t)\,=\,\left\langle \psi(\vec{r}_1,\dots,\vec{r}_N,t) \Big| \hat{\vec{P}} \Big|\psi(\vec{r}_1,\dots,\vec{r}_N,t)  \right\rangle=\,\vec{P}_0 -\hat{i}\,N\, m\,g\,t\,,
\eeq
where 
\beq
\hat{\vec{P}}\,=\,\sum_{\alpha=x,y,z}\sum_{j=1}^N \vec{e}_\alpha \, \hat{p}_{\alpha_j}
\eeq
represents the total momentum of the system, written in terms of the unit
vectors $\vec{e}_x=\hat{i}$, $\vec{e}_y=\hat{j}$ and $\vec{e}_z=\hat{k}$, while $\vec{P}_0$ is the initial $t=0$ total momentum:
\beq
\vec{P_0}\,=\,\sum_{\alpha=x,y,z}\sum_{j=1}^N \vec{e}_\alpha \, p_{\alpha 0}^{(j)}\,.
\eeq
For the total energy of the system, one can compute
\beq
E(t)\,=\,\left\langle \hat{H}\right\rangle(t)\,=\,\left\langle \psi(\vec{r}_1,\dots,\vec{r}_N,t) \Bigg| \frac{1}{2m}\hat{\vec{P}}^2 +\sum_{j<k}\hat{V}(|r_k - r_j|) +m\,g\sum_{j=1}^N \hat{x}_j \Bigg|\psi(\vec{r}_1,\dots,\vec{r}_N,t)\right\rangle\,,
\eeq
and using the above results, after an elementary but lengthy calculation, one obtains that the energy is conserved during the motion, as expected. 

Using the many-body wavefunction in Eq.~(\ref{sol_Nbodies}), we are also able to write the one-body density matrix of the falling system in terms of the
non-falling system. This is interesting because the correlation properties of many-body systems are encoded in one- and many-body density matrices, and that off-diagonal long-range order can be read out at vanishing and finite temperature
  from the one-body density matrix \cite{Pitaevskii16,Colcelli2020}. 

The one-body density matrix is defined as \cite{Pitaevskii16}:
\beq
\label{obdm_time_def}
\rho(\vec{r},\vec{r}',t)\,=\,N\,\int d\vec{r}_2\dots d\vec{r}_N\,\psi^*(\vec{r},\vec{r}_2,\dots,\vec{r}_N,t) \,\psi(\vec{r}',\vec{r}_2,\dots,\vec{r}_N,t)\,.
\eeq
Therefore using Eq.~(\ref{sol_Nbodies}) we can rewrite the density matrix as:
\beq
\rho(\vec{r},\vec{r'},t)\,=\,N\,e^{i\left[\theta(x',t)-\theta(x,t)\right]}\,\int d\boldsymbol{\varrho}_2\dots d\boldsymbol{\varrho}_N\,\eta^*(\boldsymbol{\varrho},\boldsymbol{\varrho}_2,\dots,\boldsymbol{\varrho}_N,t) \,\eta(\boldsymbol{\varrho}',\boldsymbol{\varrho}_2,\dots,\boldsymbol{\varrho}_N,t)\,,
\eeq
since $d\vec{r}_j = d\boldsymbol{\varrho}_j$ for every $j$, while
$\boldsymbol{\varrho}(t) = \vec{r} - \xi(t)$, $\boldsymbol{\varrho}'(t) = \vec{r}' - \xi(t)$,
and $x$ and $x'$ are the $x$-components of $\vec{r}$ and $\vec{r}'$ respectively. So finally:
\beq
\label{obdm_time_eta}
\rho(\vec{r},\vec{r'},t)\,=\,e^{i\left[\theta(x',t)-\theta(x,t)\right]}\,\rho_{\rm free}(\boldsymbol{\varrho},\boldsymbol{\varrho}',t)\,,
\eeq
where $\rho_{\rm free}(\boldsymbol{\varrho},\boldsymbol{\varrho}',t)$ is defined in terms of $\eta$, the solution of the Schr\"odinger equation without gravitational field.

For a translationally invariant system, the above equation may be further simplified
by writing everything in terms of the relative coordinate $\vec{R} \equiv \vec{r}-\vec{r}'$.
In this case, since $\vec{R}= \boldsymbol{\varrho}-\boldsymbol{\varrho}'$,
then Eq.~(\ref{obdm_time_eta}) may be rewritten as:
\beq
\label{obdm_transl_inv}
\rho(\vec{R},t)\,=\,e^{i \,m\,g\,t \,X/\hbar}\,\rho_{\rm free}(\vec{R},t)\,,
\eeq
where Eq.~(\ref{theta_g}) has been used and $X$ is the $x$-component of the
$\vec{R}$.

We may further analyze the eigenvalues of the one-body density matrix for a
translationally invariant system. In the static case, the one-body density
matrix satisfies the eigenvalue equation \cite{Pitaevskii16}
\beq
\label{eigeneq_obdm}
\int \rho(\vec{r},\vec{r}')\,\phi_i(\vec{r}')\,d\vec{r}'\,=\,\lambda_i \,\phi_i(\vec{r})\,,
\eeq
where $\lambda_i$ is the occupation number of the $i$--th natural
orbital eigenvector $\phi_i(\vec{r})$.
Since for a translationally invariant system the natural orbitals are simply plane waves, we can relate the natural orbitals occupation numbers of the falling system to those of the
non-falling one:
\beq
\lambda_\vec{k}(t)\,=\,\lambda_{\vec{\tilde{k}}}^{\rm free}(t)\,,
\eeq
where the wavevector
$\vec{\tilde{k}}=\left(k_x+m\,g\,t/\hbar\right)\cdot \vec{e}_x+k_y\cdot \vec{e}_y+k_z\cdot \vec{e}_z$ is the correct quantum number of the system, and we have defined:
\beq
\lambda_\vec{k}^{\rm free}(t) \,=\,\int \rho_{\rm free}(\vec{R},t)\,e^{i \vec{k} \cdot\vec{R}}\,d\vec{R}\,.
\eeq
From the above relations, one may observe that there is only a
time-dependent translation of the $x$-component of the momentum wavevector which
identifies the occupation numbers of the falling system with respect to the non-falling case
with no gravitational potential.

We emphasize again that the approach of gauge transforming and then performing a translation transformation works fine even in the case of particles with different masses, such as a free falling atom or molecule.

\section{Conclusions}

In this paper we have 
  revisited the quantum description of free fall 
  in a gravitational (or electric)
  field, and have shown that it can be nicely simplified by making use of a gauge transformation
of the wavefunction. This corresponds to changing from the laboratory reference frame to the one that moves within the falling body.
In this way we pedagogically reviewed the results already presented 
  in~\cite{Nauenberg}, and extended them to
the case of a generic three-dimensional quantum many-body system
subject to a gravitational potential. 

The gauge transformation method appears to be highly versatile and easily applicable,
since the expectation values of relevant physical quantities can be related to their counterparts in the absence of gravity. In particular we have shown that the variances of the
initial wavepacket are exactly the same as if the system doesn't feel any gravitational
force at all. Other physical observables (e.g. the one-body density matrix) are also simply related to the corresponding ones in a non-falling system.  All calculations we presented require only basic knowledge of quantum mechanics, and are therefore accessible to undergraduates.  Regarding the application of the presented method to other systems, it could be pedagogically interesting to apply it to the Dirac equation in a linear potential. 

\begin{acknowledgements} 

A. C. acknowledges fruitful correspondence with Michael Jones. A.T. acknowledges discussions
with A. P. Polychronakos during the conference ``Mathematical physics of anyons and topological states of matter” in Nordita, Stockholm (March 2019). 
Both numerical and analytical exercises on the topics presented here
were done and discussed during courses the authors taught during the years,
and we acknowledge feedback and suggestions from the students of these courses.

\end{acknowledgements}


\begin{thebibliography}{XX}
\bibitem{footnote}
Notice that, despite having a similar form, this transformation is not a gauge transformation in the conventional meaning used in electromagnetic problems.
\bibitem{footnote2}
We report for convenience only the commutators on the $x$ components, but the same commutator rules will be valid also for $y$ and $z$.
\bibitem{footnote3}
Notice however that in order to evaluate them, one needs to know how to solve the Schr\"odinger equation (\ref{twobodies_schro_eta}) for that specific interacting potential.
\end{thebibliography}
\end{document}